\documentclass[aps,pre, onecolumn, tightenlines, 12pt]{revtex4-1}
\usepackage{graphicx}
\usepackage{dcolumn}
\usepackage{comment}
\usepackage{amsmath}

\begin{document}

\title{Entanglement of pseudo-Hermitian random states}

\author{C. A. Goulart}
\affiliation{Inst\'{\i}tuto de F\'{\i}sica, Universidade de S\~{a}o Paulo\\
	Caixa Postal 66318, 05314-970 S\~{a}o Paulo, S.P., Brazil}
	\author{M. P. Pato}
\affiliation{Inst\'{\i}tuto de F\'{\i}sica, Universidade de S\~{a}o Paulo\\
	Caixa Postal 66318, 05314-970 S\~{a}o Paulo, S.P., Brazil}

\begin{abstract}
In a recent paper (A. Fring and T. Frith, Phys. Rev A {\bf 100}, 101102 (2019)), Dyson's scheme to deal with density 
matrix of non-Hermitian Hamiltonians has been used to investigate the entanglement of states of a {\it PT}-symmetric bosonic system. 
They found that von Neumann entropy can show a different behavior in the broken and unbroken regime. We show that their results can be 
recast in terms of an abstract model of pseudo-Hermitian random matrices. It is found however that, although the formalism is practically
the same, the entanglement is not of Fock states but of Bell states.  
\end{abstract}

\maketitle

\section{Introduction}

The importance of the findings of the authors of Ref. \cite{Fring2019} lies in the consequences it has to the question of time evolution of non-Hermitian Hamiltonians\cite{Mostafazadeh2007,Znojil2008}, or, more  precisely, to the evolution of their associated density matrices. It shows that by introducing an appropriate time-dependent metric, the density matrix of a non-Hermitian Hamiltonian can be linked to one of a Hermitian one by a similarity transformation. Therefore, both share the same von Neumamm entropy. From a practical point of view, Dyson's formula is an equation of motion that connects the  metric and the Hamiltonians that, in principle, can be solved as shown in \cite{Fring2019}. 

The non-Hermiticity addressed in Refs. \cite{Fring2019,Fring2017,Fring2018} is that related to the {\it PT}-symmetry. It can be shown that a complex non-Hermitian Hamiltonian invariant under the combined parity ({\it P}) and time reversal ({\it T}) transformations have eigenvalues which are real or complex conjugate. Assuming that  an operator is connected to its adjoint by a similarity transformation 
\begin{equation}
	A^{\dagger}=\eta A\eta^{-1},  \label{11}
\end{equation}
in which $\eta$ is a Hermitian operator, then its eigenvalues are real or complex conjugate. Operators satisfying this condition have been defined as belonging to the class of pseudo-Hermitian operators\cite{Mostafazadeh2002a,Mostafazadeh2002b,Mostafazadeh2002c}. This follows from the fact that using the operator $\eta$ as a metric, the internal product can be redefined such that quantum mechanics relations can be extended to the case of ${\it PT}$ symmetric Hamiltonians\cite{Bender1998,Bender1999,Bender2002,Bender2007}.

Since the beginning of the studies of ${\it PT}$-symmetric systems there was an interest in investigating random matrix ensemble to model properties of this kind of Hamiltonians. This comes naturally as time reversal symmetry plays an important role in RMT. Several ensembles already have been proposed\cite{Bohigas2013,Marinello2016a,Jain2006,Srivastava2012} but here we focus on the recently introduced ensemble of pseudo-Hermitian random matrices\cite{Marinello2016b,Marinello2017,Marinello2018,Marinello2019}. However, the matrices in these references have elements which are Gaussian distributed while constructing our model, naturally we were lead to the so-called Wishart matrices in which elements are Chi-squared distributed.  In the context of the application  of random matrix ensemble to entanglement, the Wishart ensemble
has been used to model the Schmidt eigenvalues\cite{Majumdar2008,Nadal2010,Nadal2011} of bipartite systems. In our model, however, the Wishart ensemble enters in a different way as is described below.

Another aspect of the model that we want to highlight is that its structure is analogous to that of the well-established quantum gates \cite{Deutsch1989}. In light of the recent interest in the application of random matrices in entanglement phenomena \cite{Majumdar2015,Pato2013,Vivo2016}, the task of finding Bell state \cite{Bell1964, Nielsen2009} equivalents for the matrices of this ensemble and studying their time evolution provides valuable insight into the physical nature and potential applications of such ensembles.

\section{The model}

A random model of Gaussian matrices can be introduced using as tools standard projector operators $P$ and $Q$
defined as\cite{Hussein1993}  
\begin{equation}
	P=\sum_{i=1}^{M} \left|i\right>\left<i\right| \;\;\mbox{and}\;\; Q=\sum_{j=M+1}^{N} \left|j\right>\left<j\right|
\end{equation}
with which a rectangular block of dimension $M\times (N-M$) with $N\ge 2M$ can be singled out from 
a Gaussian matrix $H$ of the Random Matrix Theory constructing with it the $N\times N$ matrix 

\begin{equation}
W = PHQ = \left(\begin{array}{c | c}
	0_{\mathbf{M}\times\mathbf{M}} & H_{\mathbf{M}\times\mathbf{N-M}} \\\hline
	0_{\mathbf{N-M}\times\mathbf{M}} & 0_{\mathbf{N-M}\times\mathbf{N-M}} \\
\end{array}\right),
\end{equation}
where $0_{\mathbf{a}\times\mathbf{b}}$ is an order $a \times b$ rectangular matrix containing all zeros, 
and $H_{\mathbf{a}\times\mathbf{b}}$ is the order $a \times b$ rectangular block from the matrix $H$. 

Combining $W$ with its adjoint, $W^{\dagger}=QHP$, three new Hermitian traceless matrices $R, S$ and $T$ 
can be defined as 
\begin{equation}
	R=W+W^{\dagger},\;\; S=-i(W-W^{\dagger}),\;\; \mbox{and} \;\; T=WW^{\dagger}-W^{\dagger}W
\end{equation}
Furthermore, by adding to these matrices, the matrix
\begin{equation}
U=WW^{\dagger}+W^{\dagger}W,
\end{equation} 
it is found that  the commutation relations 

\begin{equation}
	[R,S]=2iT,
\end{equation}  
\begin{equation}
	[S,T]=2iRU
\end{equation}  
and
\begin{equation}
	[T,R]=2iSU.
\end{equation}  
are satisfied. As $U$ commutes with the others and $R^2+S^2+T^2=2U+U^2,$ we can say that  it is a generalized 
Casimir operator. 
It is notable that these matrices have the structure of typical quantum gates \cite{Deutsch1989, Nielsen2009}. Namely, 
 $R$, $S$ and $T$ have the structure of Pauli $-X$, $-Y$ and $-Z$ gates, respectively. 

As $H$ have Gaussian elements, $WW^{\dagger}$ and $W^{\dagger}W$ are Wishart-Laguerre matrices of sizes $M\times M$ and $(N-M)\times(N-M),$ respectively. 
As a consequence, the eigenstates of $U$ are made of the decoupled eigenstates of its separated blocks. Let us denote by $\left|x_k\right>$ with $k=1,2..M$ 
and $\left|y_l\right>$  with $l=1,2,...,N-M$ the separated orthonormal eigenstates of this bipartite system, namely, we have
\begin{equation}
U\left|x_{k}\right>=WW^{\dagger}\left|x_{k}\right>=x_{k}\left|x_k\right>
\end{equation}
and
\begin{equation}
U\left|y_{l}\right>=W^{\dagger} W\left|y_{l}\right>=y_{l}\left|y_l\right>
\end{equation}
with $\left<y_{l}\right.\left|x_{k}\right>=0$.

As a matter of fact, more can be said about the $\left|y_{l}\right>$ states. Indeed, by multiplying the eigenvalue equation of $\left|x_{k}\right>$ by $W^{\dagger},$ that is
\begin{equation}
W^{\dagger}(WW^{\dagger})\left|x_{k}\right>=(W^{\dagger}W)W^{\dagger}\left|x_{k}\right>=x_{k}W^{\dagger}\left|x_k\right>
\end{equation}
we find that $W^{\dagger}|x_{k}>$ is an eigenvector of $W^{\dagger}W$ with the same eigenvalue. 
Moreover, as $\mbox {tr} ( U)=\mbox {tr}(WW^{\dagger})+\mbox {tr} (W^{\dagger}W)=2\mbox {tr}(WW^{\dagger}),$ the fact that the two operators 
share the same set of $M$ eigenvalues implies that all the other eigenvalues of $W^{\dagger}W$ are zero. 
For simplicity, we shall concern ourselves with the case in which $M \leq N/2$, but the results are easily extendable for the $M > N/2$ and quantitatively analogous.
We can, therefore, assume the first $M$ normalized eigenvectors of the operator  $W^{\dagger}W$ to be given by 
\begin{equation}
\left|y_{l}\right>=\frac{W^{\dagger}}{\sqrt{x_{l}}}\left|x_{l}\right>   \mbox{  with  } l=1,..,M 
\end{equation}

Taking into account the above bases, we consider diagonalized and reduced operators given by
\begin{equation}
\hat{U}=\sum_{k=1}^{M}\left|x_{k}\right>x_{k}\left<x_{k}\right| +\sum_{l=1}^{M}\left|y_{l}\right>x_{l}\left<y_{l}\right|
\end{equation} 

\begin{equation}
\hat{R}=\sum_{k=1}^{M}\left|x_{k}\right>\sqrt{x_{k}}\left<y_{k}\right| +\sum_{l=1}^{M}\left|y_{l}\right>\sqrt{y_{l}}\left<x_{l}\right|
\end{equation}

\begin{equation}
\hat{S}=-i\sum_{k=1}^{M}\left|x_{k}\right>x_{k}\left<y_{k}\right| +i\sum_{l=1}^{M}\left|y_{l}\right>x_{l}\left<x_{l}\right|
\end{equation} 
\begin{equation}
\hat{T}=\sum_{k=1}^{M}\left|x_{k}\right>x_{k}\left<x_{k}\right| - \sum_{l=1}^{M}\left|y_{l}\right>x_{l}\left<y_{l}\right|
\end{equation} 
that act in the space of dimension $M^2$. We remark that this result is consistent with the Schmidt decomposition of a bipartite system
\cite{Bengtsson2006}.  Accordingly,  $\hat R,$  $\hat S$ and $\hat T$ have eigenvalues $\pm \sqrt{x_{k}}$ with normalized eigenvectors

\begin{equation}
\left|X^{\pm}_{k} \right>= \frac{1}{\sqrt{2}}\left(\left|x_{k}\right>  \pm  \left|y_{k}\right>\right) ,
\end{equation}

\begin{equation}
\left|Y^{\pm}_{k} \right>=\frac{1}{\sqrt{2}}\left(\left|x_{k}\right>  \pm  i\left|y_{k}\right>\right)  \label{737}
\end{equation}
and 

\begin{equation}
\left|Z^{+}_{k}\right>=\left|x_{k}\right> \mbox{   and   }\left|Z^{-}_{k}\right>=\left|y_{k}\right> \label{735}
\end{equation}
respectively.

This also implies that those interactions may be seen as analogues to the effect of quantum gates on qu-$d$it states, where the the digit 
is determined by $M$\cite{Daoud2018}. Namely, the corresponding Bell state is composed of two qu-$M$its, corresponding to the dimension of the images 
of $WW^\dagger$ and $W^\dagger W$.

Moreover, as $\hat U$  is positive, it can be used to perform the polar decomposition\cite{Bengtsson2006} of our matrices  such that  a vector 
operator can be defined as ${\bf g}=(\hat{U}^{-1/2} \hat{R},\hat{U}^{-1/2} \hat{S},\hat{U}^{-1} \hat{T})$. 
Then the above commutation relations can be rewritten as 

\begin{equation}
[g_i, g_j] = 2 i \varepsilon_{i j k} g_k , 
\end{equation}
which together with the  anti-commutation relations
\begin{equation}
\{ g_{i},g_{j}\} = 2 \delta_{i j} \textbf{1},
\end{equation}

where $\textbf{1}$ is the identity matrix, confirm that the set of unitary matrices $g_{i}$ have the same structure of the Pauli matrices. 
They are therefore a $SU(2)$ 
representation. In fact, for $M=1$ and $N=2,$
the three $g_{i}$ matrices indeed coincide with the Pauli matrices.  For $M>1$, the $g_i$ are extensions of the Pauli matrices
in which the zeros and the ones have dimension $M$, namely they are $\textbf{0}_{M\times M}$ and $\textbf{1}_{M\times M}$.
We also notice that the Baker-Campbell-Hausdorff formula provides the useful expansion

\begin{equation}
\exp( ag_{i})g_{j}\exp(-a g_{i})=g_{j}\cosh(2a) +\frac{1}{2} [g_{i}, g_{j}]\sinh(2a) \mbox{   with  } i\ne j .
\end{equation}

Using the Bloch vector ${\bf u}=(\sin\theta\cos\phi,\sin\theta\sin\phi,\cos\theta),$ the operator

\begin{equation}
	\begin{split}
		\frac{1+{\bf u}.{\bf g}}{2} &={\hat  U}^{-1} \left(\frac{\hat U+\hat T}{2} \cos^{2} \frac{\theta}{2}+\frac{\hat U-\hat T}{2} \sin^{2} \frac{\theta}{2}\right)\\ 
		&+\sqrt{\hat U}^{-1}\left(\frac{\hat R+i\hat S}{2}\frac{\sin\theta}{2} e^{-i\phi}+\frac{\hat R-i\hat S}{2}\frac{\sin\theta}{2} e^{+i\phi}\right)
	\end{split}	
\end{equation}
is constructed that has eigenvalue one and eigenvectors 
\begin{equation}
	\cos\frac{\theta}{2}\left|x_{k}\right> + \sin\frac{\theta}{2} e^{i\phi}  \left|y_{k}\right> \label{83}
\end{equation}
and 
\begin{equation}
	\sin\frac{\theta}{2}\left|y_{l}\right> + \cos\frac{\theta}{2} e^{-i\phi} \left|x_{l}\right> .
\end{equation}
Therefore, it is a projector and the eigenvectors are pure states in the Bloch sphere. 

The above operator is a Hermitian linear combination of the generators. By making imaginary one of the coefficients, non-Hermitian  linear combination of these 
matrices can be introduced. 
Once this is done, Dyson's formula can then be used to eliminate the non-Hermitian term in such a way that an associated Hermitian Hamiltonian is produced.

Before doing this, it is convenient to recall some properties of density matrices. Let a density matrix of a non-Hermitian matrix $A$ with a set of $N$   states $\left|\Psi_{i}\right>$ be written as
\begin{equation}
\rho_{A}=\sum_{i=1}^{N} p_{i}\left|\Psi_{i}\right>\left<\Psi_{i}\right|\mu^{\dagger}\mu ,
\end{equation}  
where $\sum_{i=1}^{N}p_{i}=1$ and $\mu^{\dagger}\mu$ is a metric. By multiplying by $\mu$ from the left and, by its inverse $\mu^{-1}$ from the right, we get 
\begin{equation}
\mu\rho_{A}\mu^{-1}=\sum_{i=1}^{N} p_{i}\mu\left|\Psi_{i}\right><\Psi_{i}|\mu^{\dagger},
\end{equation}  
which can be interpreted as the density matrix $\rho_{h}=\mu\rho_{A}\mu^{-1}$ of a Hermitian matrix $h$. By construction both density matrices share the same set of eigenvalues. It can be shown \cite{Fring2016a,Fring2016b} that the Hermitian $h$ is connected to the non-Hermitian $A$ by the relation
\begin{equation}
h=\mu A\mu^{-1} + i\delta_{t}(\mu)\mu^{-1}\label{dys1},
\end{equation}
which we review in the Appendix. This is Dyson's formula that provides a way to determine the Hermitian Hamiltonian $h$. It is important to remark that the above derivation also holds for 
a time-dependent  Hamiltonians and metrics \cite{Fring2016a,*Fring2016b}.

In the sequel, this procedure is performed for two pseudo-Hermitian Hamiltonians $A_{1}$ and $A_{2}$ which are linear combination of our matrices.

\subsection{$A_1$ matrix}

We start by considering the case of the matrix
\begin{equation}
	A_{1}=\hat{U}+b\hat{R}+ic\hat{S}
\end{equation}
wihere $b$ and $c$ are real non-negative coefficients. After some straightforward algebra, it can be verified that $A_{1}$ has eigenvalues 
\begin{equation}
	x_{k}\pm \sqrt{b^2 -c^2}\sqrt{x_k} \mbox{   with   } k=1,2,...,M
\end{equation}
with eigenvectors
\begin{equation}
	\left|x_{k}\right> \pm \sqrt{\frac{b -c}{b+c}}  \left|y_{k}\right>.
\end{equation}
As the eigenvalues are real, if $b\ge c$, and complex conjugate, if $b<c$, $A_{1}$ is pseudo-Hermitian Hamiltonian. Inspired by Ref \cite{Fring2019}, we make the ansatz
\begin{equation}
	\mu=\exp(\beta \hat{S})\exp(\alpha \hat{T})
\end{equation}
that replaced in Eq. (\ref{dys1}) allow us to derive, with the help  of the above relations, that
\begin{equation}
	\begin{split}
		h=\hat{U} &+ \left[R\cosh(2\beta\sqrt{\hat{U}})-\frac{i\hat{T}}{\sqrt{\hat{U}}} \sinh(2\beta\sqrt{\hat{U}})\right]\left[b\cosh(2\alpha \hat{U}) + c\sinh(2\alpha \hat{U})\right]  \\
		&+iS\left[b\sinh(2\alpha \hat{U}) + c\cosh(2\alpha \hat{U})\right] -i\dot\beta \hat{S} \\
		&+\left[\hat{R}\sinh(2\beta\sqrt{\hat{U}})-\frac{i\hat{T}}{\sqrt{\hat{U}}} \cosh(2\beta\sqrt{\hat{U}})\right]\sqrt{\hat{U}}\dot\alpha.
	\end{split}
\end{equation}
By imposing that the parameters $\alpha$ and $\beta$ satisfy the differential equations
\begin{equation}
	\dot\alpha=-\frac{\tanh(2\beta \sqrt{\hat{U}})}{\sqrt{\hat{U}}} \left[b\cosh(2\alpha \hat{U}) + c\sinh(2\alpha \hat{U})\right] \label{Fring1}
\end{equation}
and
\begin{equation}
	\dot\beta=b\sinh(2\alpha \hat{U}) + c\cosh(2\alpha \hat{U}) \label{Fring2},
\end{equation}
the non-Hermitian terms are removed and the Hermitian matrix 
\begin{equation}
	h=\hat{U}+\frac{b\cosh(2\alpha \hat{U}) + c\sinh(2\alpha \hat{U})}{\cosh(2\beta\sqrt{\hat{U})}}\hat{R} =\hat{U} +\nu(t) \hat{R} . \label{R5}
\end{equation}
is obtained. 

Although these equations show great similarity to those of  \cite{Fring2019}, an important difference lies in the fact that for them
$\alpha$ and $\beta$ are numbers while for us they are matrices that are functions of $\hat{U}$.
Nevertheless, relying in the commutativity of $U$ with the other matrices, we assume
that $U$ can be treated as a number, such that our equations can be solved  following the same steps. Thus, by inverting (\ref{Fring2}) we obtain
\begin{equation}
	\tanh(2\alpha \hat{U})=\frac{-bc+\dot\beta\sqrt{b^2-c^2+\dot\beta^2}}{b^2+\dot\beta^2}. \label{67}
\end{equation}
Next, in order to decouple Eqs. (\ref{Fring1}) and (\ref{Fring2}), we first combine them to get
\begin{equation}
	\dot\alpha=-\frac{1}{\sqrt{\hat{U}}}\tanh(2\beta\sqrt{\hat{U}})\sqrt{b^2 +\dot\beta^2-c^2},  \label{alpha}
\end{equation}
then, by taking the derivative of (\ref{Fring2}), we obtain
\begin{equation}
	\ddot{\beta}=-2\sqrt{\hat{U}}\tanh(2\beta\sqrt{\hat{U}})(\dot\beta^2+b^2-c^2)
\end{equation}
such that,  if a new variable $\sigma=\sinh(2\beta\sqrt{\hat U})$ is defined, it is found that it satisfies a harmonic oscillator equation that can be solved as  
\begin{equation}
	\sinh(2\beta\sqrt{\hat{U}})=\frac{C_{1}}{\sqrt{b^2-c^2}}\sin\left[2\sqrt{\hat{U}(b^2 - c^2)}(t+C_2)\right] .
\end{equation}
Once, the dependence with time of the parameter $\beta$ is obtained, we can replace it in Eq. (\ref{alpha}) to have
\begin{equation}
	\sqrt{\hat{U}}\dot\alpha=-\frac{\sqrt{(b^2-c^2)(C_{1}^2+b^2 -c^2)}C_{1}\sin\left[2\sqrt{\hat{U}(b^2 - c^2)}(t+C_2)\right] }
	{b^2-c^2+C_{1}^2\sin^{2}\left[2\sqrt{\hat{U}(b^2 - c^2)}(t+C_2)\right] }
\end{equation}
that can be solved as
\begin{equation}
	2\hat{U}\alpha=\mbox{atanh}\left(\frac{C_{1}\cos\left[2\sqrt{\hat{U}(b^2 - c^2)}(t+C_{2})\right]}
	{\sqrt{C_{1}^2+b^2-c^2}}\right) +B,
\end{equation}
where $B$ is an integration constant which with the value $B=-\mbox{atanh}(c/b)$ yields
the same expression

\begin{equation}
\exp(4\alpha \hat{U})=\left(\frac{b-c}{b+c}\right)\frac{\sqrt{C_{1}^2+b^2-c^2}+C_{1}\cos\left[2\sqrt{\hat{U}(b^2 - c^2)}(t+C_2)\right]}
{\sqrt{C_{1}^2+b^2-c^2}-C_{1}\cos\left[2\sqrt{\hat{U}(b^2 - c^2)}(t+C_2)\right]},
\end{equation} 
in \cite{Fring2019} (we remark that the above value of the integration constant is necessary in order to Eq. (\ref{67}) be satisfied). From it, we obtain
\begin{equation}
\nu(t)=\frac{(b^2-c^2)\sqrt{C_{1}^2+b^{2}-c^2}}{C_{1}^2+b^2-c^2- 
C_{1}^{2}\cos^{2}\left[2\sqrt{\hat{U}(b^2 - c^2)}(t+C_2)\right]} .
\end{equation}
and
\begin{equation}
	\nu_{I}=\int^{t}\nu(t^{\prime})dt^{\prime}=\frac{1}{2\sqrt{U}}\arctan\left(\frac{\sqrt{C_{1}^2+b^2 -c^2}}{\sqrt{b^2 -c^2}}
	\tan[2\sqrt{\hat{U}(b^2-c^2)}(t+C_{2})]\right) =\frac{\gamma(\hat{U})}{\sqrt{\hat{U}}}. \label{231}
\end{equation}
such that we now have the unitary operator $\exp\left(-i\hat Ut-i\gamma(\hat U)\frac{\hat R}{\sqrt{\hat U}}\right)$ to make the time evolution 
of initial states.

Let's starting evolving just the single state given by Eq. (\ref{83}). Applying the term $\exp\left(-i\hat U\right)$ it
just produces an overall phase and, for the other term, we have 

\begin{equation}
\gamma(\hat U)\frac{\hat R}{\sqrt{\hat U}}=\sum_{k=1}^{M}\left|x_{m}\right>\gamma(x_{m})\left<y_{k}\right| +
\sum_{l=1}^{M}\left|y_{l}\right>\gamma(y_{l})\left<x_{l}\right|
\end{equation}
such that, after some algebra, we obtain

\begin{equation}
e^{-i\hat{U}-i\frac{\gamma(\hat{U})R}{\sqrt{\hat{U}}}}\left(\cos\frac{\theta}{2}\left|x_{k}\right> +e^{i\phi}\sin\frac{\theta}{2}  \left|y_{k}\right>\right)=e^{-ix_{k}t}
\left[\cos(\frac{\theta}{2}-\gamma_{k})\left|x_{k}\right> +e^{i\phi}\sin(\frac{\theta}{2}-\gamma_{k})  \left|y_{k}\right>\right].
\end{equation} 
The above equation shows that the generator produces a rotation and the evolved density matrix is then

\begin{equation}
	\begin{split}
		\rho=&\cos^{2}(\frac{\theta}{2}-\gamma_{k}) \left|x_{k}\right>\left<x_{k}\right| + \sin^{2}(\frac{\theta}{2}-\gamma_{k}) \left|y_{k}\right>\left<y_{k}\right| \\
		&+\cos(\frac{\theta}{2}-\gamma_{k}) \sin(\frac{\theta}{2}-\gamma_{k}) \left(e^{i\phi}\left|x_{k}\right>\left<y_{k}\right|
		 - e^{-\phi}\left|y_{k}\right>\left<y_{k}\right| \right),
	\end{split}
\end{equation}
which is a pure state. If the partial trace is taken or, in physical terms, if the interference terms are removed, the states become mixed in which 
$ \cos^{2}(\frac{\theta}{2}-\gamma_{k})$ and $\sin^{2}(\frac{\theta}{2}-\gamma_{k}) $ 
are the probabilities  of finding, in a measurement,  the system to be in $PHQHP$ or in $QHPHQ$ parts, respectively. 
In this case, if $b >c,$ the probabilities oscillate, while, for $b<c,$ asymptotically, when $t$ goes to infinity, $\gamma_{k}$ goes to 
$\gamma_{\infty}=\frac{1}{2}\arctan\sqrt{\frac{C_{1}^2 +b^2 -c^2}{c^2 -b^2}}$
and the probabilities reach fixed values.    

Turning now to entanglement, one would expect to be natural to study the bipartite division $PHQHP$ and $QHPHQ,$ but, taking into account that 
the eigenstates of our matrices are qubits, we work with them instead. Considering the present case of the generator $\hat R,$ Eq. (\ref{735}), 
we use pairs $(m,n)$ of its eigenstates to construct the Bell states

\begin{equation}
\left|\Phi^{\pm}_{R}(0)\right>= \frac{1}{\sqrt{2}}\left(\left|X_{m}^{+}\right>\left|X_{n}^{+}\right>  \pm \left|X_{m}^{-}\right>\left|X_{n}^{-}\right> \right) ,
\end{equation}
with which an initial state 

\begin{equation}
\left|\chi(0)\right>=\cos\frac{\theta}{2}\left|\Phi^{+}_{R}(0)\right>  +  \sin\frac{\theta}{2} \left|\Phi^{-}_{R}(0)\right>,
\end{equation}
is defined. In terms of the qubit states, this initial state can be spanned as 

\begin{equation}
\left|\chi(0)\right>=\Lambda_{11}\left|X_{m}^{+}\right>\left|X_{n}^{+}\right>  + \Lambda_{12}\left|X_{m}^{+}\right>\left|X_{n}^{-}\right> 
+\Lambda_{21}\left|X_{m}^{-}\right>\left|X_{n}^{+}\right>  + \Lambda_{22}\left|X_{m}^{-}\right>\left|X_{n}^{-}\right> 
\end{equation}
and a density matrix  $\rho_{ij;kl}=\Lambda_{ij}\Lambda_{kl}$ where

\begin{equation}
\Lambda (0) =\frac{1}{\sqrt{2}}\left(\begin{array}{c  c}
	    \cos\frac{\theta}{2} +\sin\frac{\theta}{2} & 0  \\
	    0  & \cos\frac{\theta}{2}- \sin\frac{\theta}{2}  \\
\end{array}\right)
\end{equation}
follows. Then, by taking partial trace, the $n$-states are removed and the reduced matrix density

\begin{equation}
\rho_{ik}^{m}(0)=\sum_{l=1}^{2}\Lambda_{il}\Lambda_{kl} = \frac{1}{2}\left(\begin{array}{c  c}
	1+\sin\theta & 0 \\
	0 & 1- \sin\theta \\
\end{array}\right)
\end{equation}
is obtained. We observe that for $\theta$ equal to zero or $\pi,$ as the initial state turns out to be one of the Bell states, the
entanglement is maximum. On the other hand, for  $\theta=\frac{\pi}{2},$ the initial state is a pure state. 

Evolving now the above initial state, we observe that, as

\begin{equation}
\frac{\gamma(\hat{U})\hat R}{\sqrt{\hat{U}}} \left|\Phi^{\pm}_{R}(0)\right>=
(\gamma_{m}+\gamma_{n})\left|\Phi^{\mp}_{R}(0)\right> ,
\end{equation}
it is deduced that

\begin{equation}
e^{-i\frac{\gamma(\hat{U})R}{\sqrt{\hat{U}}}} \left|\Phi^{\pm}_{R}\right>=\left|\Phi^{\pm}_{mn}(t)\right>=\cos\Delta\left|\Phi^{\pm}_{R}(0)\right> -
i \sin\Delta \left|\Phi^{\mp}_{R}(0)\right>,
\end{equation}
where $\Delta=\gamma_{m}+\gamma_{n}.$ From this result, it follows that

\begin{equation}
\left|\chi(t)\right>=(\cos\frac{\theta}{2}\cos\Delta-i\sin\frac{\theta}{2}\sin\Delta)\left|\Phi^{+}_{R}(t)\right>  +
 (\sin\frac{\theta}{2}\cos\Delta-i\cos\frac{\theta}{2}\sin\Delta )\left|\Phi^{-}_{R}(t)\right>,
\end{equation}
or, in terms of the eigenstates, 

\begin{multline}
\left|\chi(t)\right>=\frac{1}{\sqrt{2}}[(\cos\frac{\theta}{2}+\sin\frac{\theta}{2})\cos\Delta-i(\sin\frac{\theta}{2}+\cos\frac{\theta}{2})\sin\Delta]\left|X^{+}_{m}\right> \left|X^{+}_{n}\right> \\
+\frac{1}{\sqrt{2}} [(\cos\frac{\theta}{2}-\sin\frac{\theta}{2})\cos\Delta-i(\sin\frac{\theta}{2}-\cos\frac{\theta}{2})\sin\Delta]\left|X^{-}_{m}\right> \left|X^{-}_{n}\right>.
\end{multline}
Taking then the partial trace, the time dependent reduced density matrix

\begin{equation}
\rho_{ik}^{m}(t) =\frac{1}{2} \left(\begin{array}{c  c}
	1 + \sin\theta\cos 2\Delta & 0 \\
	 0 & 1 -  \sin\theta\cos 2\Delta  \\
\end{array}\right)
\end{equation}
is obtained and the von Neumann entropy

\begin{equation}
S=-\lambda_{1}\log \lambda_{1}- \lambda_{2}\log \lambda_{2}
\end{equation}
is calculated with

\begin{equation}
	\lambda_{1}= \frac{1}{2}\left(1 + \sin\theta\cos 2\Delta\right) 
\end{equation}
and
\begin{equation}
	\lambda_{2}= \frac{1}{2}\left(1 - \sin\theta\cos 2\Delta\right) .
\end{equation}
We remark that the entangled qubits are chiral states of the matrix $R$ and also of the pseudo-Hermitian matrix $A_1.$ 
Their entanglement is done in a $++$ and  $--$ way such that, if the state $m$ is in the positive (negative) state then the $n$ is in the positive (negative).

In Fig.1 and Fig. 2, it is shown the time evolution of the von Neumann entropy in the regimes of real, $b>c,$ and complex conjugate eigenvalues, $b<c,$ respectively.
In both cases, $N=6$ and $M=2$ such that there are only the ground state and the first excited state besides, as $\theta=\frac{\pi}{2}$, the initial is a pure state. 
In Fig. 1, the calculation is performed with two sample matrices and the effect of the randomness 
of the eigenvalues is exhibited. This randomness is  also present in the fact that the oscillations do not show a constant period. As a matter of fact, for larger matrices, 
that is $N\gg M$ the randomness is suppressed. 

In Fig. 2,  the "eternal life of the entropy" effect, as the authors of Ref. \cite{Fring2019} called it, is shown. Starting from its value at the pure state, the entropy reaches the
value of maximum entanglement and then decays to a constant value.

\begin{figure}[ht]
\centerline{\includegraphics*[angle=0,width=0.50\textwidth]{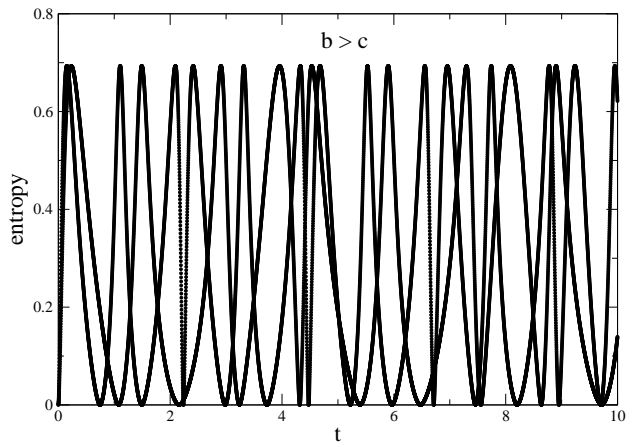}}
\caption{With $\theta=\frac{\pi}{2}$, von Neumann entropy in function of time for two random matrices with the values $C_{1}=2,$ $b=1.2$ and $c=1.$}
\end{figure}

\begin{figure}[ht]
\centerline{\includegraphics*[angle=0,width=0.50\textwidth]{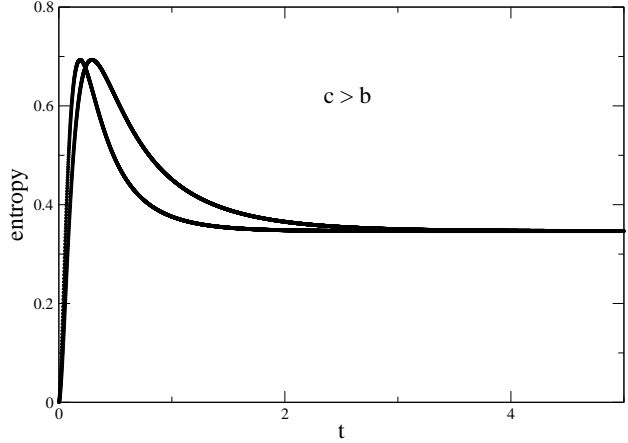}}
\caption{With $\theta=\frac{\pi}{2}$, von Neumann entropy in function of time for two random matrices with the values $C_{1}=2,$ $b=1.0$ and $c=1.2$.}
\end{figure}

\subsection{$A_2$ matrix}

We now consider the matrix
\begin{equation} 
	A_{2}=\hat U+b\frac{\hat T}{\sqrt{\hat U}} - ic\hat S,
\end{equation}
in which, as before, $b$ and $c$ real non-negative coefficients. It can be again verified that $A_{2}$ has eigenvalues 
\begin{equation}
x_{k}\pm \sqrt{b^2 -c^2}\sqrt{x_k} \mbox{   with   } k=1,2,...,M
\end{equation}
with unnormalized eigenvectors
\begin{equation}
\left(b\pm\sqrt{b^2-c^2}\right)\left|x_{k}\right> +c \left|y_{k}\right>.
\end{equation}
The eigenvalues are real, if $b\ge c$, and complex conjugate, if $b<c$, $A_{2}$ is pseudo-Hermitian with respect to the metric $P-Q$.

Replacing  the ansatz
\begin{equation}
\mu=\exp(\zeta \hat S)\exp(-\delta \hat T)
\end{equation}
in Eq. (\ref{dys1}) we find that if  the time dependent parameters $\zeta$ and $\delta$ satisfy the differential equations
\begin{equation}
\dot\delta=-\frac{\tanh(2\zeta \sqrt{U})}{\sqrt{\hat U}} \left[b\cosh(2\delta \sqrt{\hat U}) + c\sinh(2\delta \sqrt{\hat U})\right] \label{Fring3}
\end{equation}
and
\begin{equation}
\dot\zeta=b\sinh(2\delta \sqrt{\hat U}) + c\cosh(2\delta \sqrt{U}) \label{Fring4},
\end{equation}
then, following the same steps employed  in the $A_{1}$ Hamiltonian, the non-Hermitian terms are removed and the Hermitian matrix 
\begin{equation}
h=\hat U+\frac{b\cosh(2\delta \sqrt{\hat U}) + c\sinh(2\delta \sqrt{\hat U})}{\cosh(2\zeta\sqrt{\hat U)}}\hat R =\hat U +\xi(t) \frac{T}{\sqrt{U}} . \label{R7}
\end{equation}
is obtained.

Evolving firstly just the single state given by Eq. (\ref{83}) with $\xi_{I}(t)$ equal to $\nu_{I}(t)$, Eq. (\ref{231}), we find the evolved state

\begin{equation}
	\begin{split}
		\left|\chi(t)\right>=&\exp\left(-iU-i\gamma \frac{\hat T}{\hat U}\right)\left(\cos\frac{\theta}{2}\left|x_{k}\right> + 
		\sin\frac{\theta}{2} e^{i\phi}  \left|y_{k}\right>\right) \\
		=&\exp(-ix_{k})\left(e^{-i\gamma_{k}}\cos\frac{\theta}{2}\left|x_{k}\right> +
		e^{i\gamma_{k}+i\phi}\sin\frac{\theta}{2}  \left|y_{k}\right>\right).
	\end{split}	
\end{equation}
and the evolved density matrix 

\begin{equation}
	\begin{split}
		\rho=\left|\chi(t)\right>\left<\chi(t)\right|=  &\cos^{2}\frac{\theta}{2} \left|x_{k}\right>\left<x_{k}\right| + \sin^{2}\frac{\theta}{2} \left|y_{k}\right>\left<y_{k}\right| \\
		&+\cos\frac{\theta}{2} \sin\frac{\theta}{2} \left( e^{-2i\gamma_{k}-i\phi} \left|x_{k}\right>\left<y_{}\right|
		 + e^{2i\gamma_{k}+i\phi}\left|y_{k}\right>\left<y_{k}\right| \right),
	\end{split}
\end{equation}
which is a pure state. The evolution produces a phase and, as a consequence, the quantum measurement is not modified.

Turning now to entanglement, we use the eigenstates of the operator $\hat T,$ Eq. (\ref{735}),  to constructing the Bell states

\begin{equation}
\left|\Phi^{\pm}_{T}(0)\right>= \frac{1}{\sqrt{2}}\left(\left|x_{m}\right>\left|x_{n}\right>  \pm \left|y_{m}\right>\left|y_{n}\right> \right) 
\end{equation}
and define the initial state 

\begin{equation}
\left|\chi(0)\right>=\cos\frac{\theta}{2} \left|\Phi^{+}_{T}(0)\right> + \sin\frac{\theta}{2} \left|\Phi^{-}_{T}(0)\right>.
\end{equation}
Using alternatively tensor product we have

\begin{equation}
\left|\Phi^{\pm}_{T}(t)\right>= e^{-i\gamma(\hat U)\frac{\hat T}{\hat U}} \otimes e^{-i\gamma(\hat U)\frac{\hat T}{\hat U}} \left|\Phi^{\pm}_{T}(0)\right>=
\cos\Delta\left|\Phi^{\pm}_{T}(0)\right> -i \sin\Delta \left|\Phi^{\mp}_{T}(0)\right>,
\end{equation}
and, after taking the partial trace, the same time dependent reduced density matrix

\begin{equation}
\rho_{ik}^{m}(t) =\frac{1}{2} \left(\begin{array}{c  c}
	1 + \sin\theta\cos 2\Delta & 0 \\
	 0 & 1 -  \sin\theta\cos 2\Delta  \\
\end{array}\right).
\end{equation}
is obtained. We have, therefore, ended up with the same von Neumann entropy of the previous case, but with a different underlying physics. Here the qubits are the eigenstates of the matrix $T$  
and also of the pseudo-Hermitian $A_2.$ The entanglement here is such that the positive and the negative values correspond to the bipartite 
$PHQHP$ and $QHPHQ$ parts of the system. If the $m$-qubit is $+,$ both states are in the first part while, if it is $-,$ both are in second part. 

\section{Conclusion}

Motivated by the fact, proved in Ref. \cite{Fring2019}, that the time evolution of the density matrix of a non-Hermitian
Hamiltonian can be connected to  one of a Hermitian Hamiltonian, we have constructed an abstract random matrix model 
to discuss this important finding. The model is made of three traceless random matrices with a structure of Pauli matrices 
whose eigenstates are qubits. A fourth matrix constituted of a special bipartite system of Wishart matrices whose parts 
share the same set of main eigenvalues plays the role of a Casimir operator. Combining then these matrices, pseudo-Hermitian
Hamiltonians are introduced by turning non-Hermitian one of the traceless matrix. Using in Dyson scheme the ansatz of 
Ref. \cite{Fring2019}, the non-Hermitian 
term is removed leaving behind a Hermitian Hamiltonian with two terms: the Casimir term and a generator term. Bell states are then 
constructed using the generator eigenstates and an appropriate superposition of them are evolved in time. In the evolution
of the von Neumenn entropy is then observed, in the complex conjugate regime of the eigenvalues, the "eternel life of the entropy"
effect as the authors  of Ref.  \cite{Fring2019} called it.
Namely, starting from its value of maximum entanglement, asymptotically, the entropy decays towards a fixed value of entanglement.
The fact that we are finding the same effect in a different physical context enhances its range of potential practical application.

\section{Acknowledgments}

The authors acknowledge fruitful discussions with Gabriel Marinello and  M. H. Y. Moussa.
M. P. P. was supported by grant 307807/2017-7 of the Brazilian agency Conselho Nacional de Desenvolvimento Científico e 
Tecnológico (CNPq) and
is a member of the Brazilian National Institute of Science and Technology-Quantum Information (INCT-IQ).  C. A. G. is supported
by a CNPq grant (Processo: 133394/2019-0).

\appendix

\section{Dyson's formula}

In this appendix, we review the derivation fo Dyson's formula. The Schr\"{o}edinger equations for $h$ is
\begin{equation}
i\delta_{t}\left(\mu\left|\Psi\right>\right)=h\left(\mu\left|\Psi\right>\right)
\end{equation}
and the evolution equation for its density matrix is
\begin{equation}
i\delta_{t}\rho_{h}=[h,\rho_{h}].
\end{equation}
Expressing  in this equation $\rho_{h}$ in terms of $\rho_{A},$ we have
\begin{equation}
i\delta_{t}\left(\mu\rho_{A}\mu^{-1}\right)=[h,\mu\rho_{A}\mu^{-1}].
\end{equation}
that leads to 
\begin{equation}
i\delta_{t}(\mu)\rho_{A}\mu^{-1}+ \mu\rho_{A}i\delta_{t}(\mu^{-1}) +
\mu i\delta_{t}(\rho_{A})\mu^{-1}=[h,\mu\rho_{A}\mu^{-1}].
\end{equation}
Multiplying from left by $\mu^{-1}$ and by $\mu$ from the right
\begin{equation}
i\mu^{-1} \delta_{t}(\mu)\rho_{A}+ \rho_{A}i\delta_{t}(\mu^{-1})\mu +
i\delta_{t}\rho_{A}=\mu^{-1} [h,\mu\rho_{A}\mu^{-1}] \mu.
\end{equation}
and using that $i\mu^{-1} \delta_{t}(\mu)=-i\delta_{t}(\mu^{-1})\mu $ we obtain
\begin{equation}
i\delta_{t}\rho_{A}=\left[\mu^{-1} h\mu-i\mu^{-1}\delta_{t}(\mu),\rho_{A} \right]
\end{equation}
such that, in order to $\rho_{A}$ to satisfy an evolution equation of the form $i\delta_{t}\rho_{A}=[A,\rho_{A}]$ the differential equation
\begin{equation}
h=\mu A\mu^{-1} + i\delta_{t}(\mu)\mu^{-1}\label{dys1:app}
\end{equation}
must be satisfied.

\bibliography{refs}

\end{document}